\documentclass{article}
\usepackage{amsmath}
\usepackage{txfonts}
 \title{Bilinearization and Casorati determinant solution to the
 non-autonomous discrete KdV equation}
\author{\textsc{Kenji Kajiwara}$^1$\thanks{E-mail address: kaji@math.kyushu-u.ac.jp}
and \textsc{Yasuhiro Ohta}$^2$\thanks{E-mail address: ohta@math.sci.kobe-u.ac.jp}}

\begin{document}
\begin{center}
 {\Large Bilinearization and Casorati determinant solution to the
 non-autonomous\\[1.5mm] discrete KdV equation}\\[4mm]
\textsc{\large Kenji Kajiwara}\\[1mm]
Faculty of Mathematics, Kyushu University,
6-10-1 Hakozaki, Fukuoka 812-8581, Japan\\[2mm]
\textsc{\large Yasuhiro Ohta}\\[1mm]
Department of Mathematics, Kobe University, Rokko, Kobe 657-8501, Japan
\end{center}
\begin{abstract}
Casorati determinant solution to the non-autonomous discrete KdV
equation is constructed by using the bilinear formalism. We present
three different bilinear formulations which have different
origins.
\end{abstract}
\section{Introduction} 
In this article, we consider the following partial difference equation
\begin{equation}
 \left(\frac{1}{a_m}+\frac{1}{b_{n+1}}\right) v_{n+1}^m
-\left(\frac{1}{a_{m+1}}+\frac{1}{b_{n}}\right) v_{n}^{m+1}
=
 \left(\frac{1}{a_m}-\frac{1}{b_{n}}\right) \frac{1}{v_{n}^{m}}
-\left(\frac{1}{a_{m+1}}-\frac{1}{b_{n+1}}\right)\frac{1}{v_{n+1}^{m+1}},
\label{eqn:ndKdV}
\end{equation}
where $m$, $n$ are the discrete independent variables, $v_n^m$ is the
dependent variable on the lattice site ($m,n$), and $a_m$, $b_n$ are
arbitrary functions of $m$ and $n$, respectively. Recently
eq.(\ref{eqn:ndKdV}) has been derived by Matsuura\cite{Matsuura} as the
equation of motion of discrete curves on the centro-affine plane.
In this context, $v_n^m$ is related to the curvature and it is essential that
$a_m$ and $b_n$ depend on $m$ and $n$,
respectively.  For analyzing the motion of discrete curve,
constructing solutions of eq.(\ref{eqn:ndKdV}) explicitly is an
interesting subject.  We
call eq.(\ref{eqn:ndKdV}) the non-autonomous discrete KdV equation, for
if $a_m$ and $b_n$ are constants, e.g. $a_m=a$, $b_n=b$,
eq.(\ref{eqn:ndKdV}) reduces to the discrete KdV
equation\cite{Hirota:dKdV,OHTI:dKP}
\begin{equation}
 \left(\frac{1}{a}+\frac{1}{b}\right) v_{n+1}^m
-\left(\frac{1}{a}+\frac{1}{b}\right) v_{n}^{m+1}
=
 \left(\frac{1}{a}-\frac{1}{b}\right) \frac{1}{v_{n}^{m}}
-\left(\frac{1}{a}-\frac{1}{b}\right)\frac{1}{v_{n+1}^{m+1}},
\label{eqn:dKdV}
\end{equation}
or
\begin{equation}
v_{n+1}^m
-v_{n}^{m+1}
=
\frac{a-b}{a+b}
 \left(\frac{1}{v_{n+1}^{m+1}}-\frac{1}{v_{n}^{m}}\right) .
\label{eqn:dKdV2}
\end{equation}
Since the constants $a$ and $b$ correspond to the lattice intervals of
$m$ and $n$, respectively, eq.(\ref{eqn:ndKdV}) can be also regarded as
the discrete KdV equation on inhomogeneous lattice.

The non-autonomous version of discrete integrable systems on
two-dimensional lattice have not been investigated well, although those on
three-dimensional lattice, such as the Hirota-Miwa (discrete KP)
equation or the
discrete two-dimensional Toda lattice equation, have been studied well
together with their
solutions\cite{KS:q2dTL,KOS:q2dTM,NTSWK:u2dTL,WTS1,WTS2}.  Let us
explain a reason taking eq.(\ref{eqn:ndKdV}) as an example. The
autonomous version eq.(\ref{eqn:dKdV}) can be transformed into so-called
the bilinear equation by suitable dependent variable transformation. The
bilinear equation is regarded as a reduction of the Hirota-Miwa
equation\cite{Hirota:DAGTE,Miwa}, which is well-known to admit various
types of exact solutions, such as soliton
solutions\cite{OHTI:dKP,Hirota:DAGTE}, rational solutions expressible in
terms of the Schur functions\cite{Miwa,Jimbo-Miwa}, or periodic
solutions that are written in terms of the Riemann theta
functions\cite{KWZ:elliptic}. Therefore one can obtain solutions to the
discrete KdV equation (\ref{eqn:dKdV}) by applying the reduction
procedure to those for the Hirota-Miwa equation. Now, the Hirota-Miwa
equation and its solutions can be generalized to non-autonomous case in
a straightforward manner.  However, it is shown that one cannot apply
the reduction procedure to the non-autonomous Hirota-Miwa equation
consistently. Moreover, eq. (\ref{eqn:ndKdV}) cannot be put into
bilinear equation by the procedure similar to the autonomous case
because of the non-autonomous property.  Therefore it was not clear how
to construct solutions to the two-dimensional non-autonomous discrete
integrable systems systematically.

In this article, we construct the Casorati determinant solution to the
non-autonomous discrete KdV equation (\ref{eqn:ndKdV}) by using the
bilinear formalism. We present three different bilinearizations: The
first one can be derived by the reduction of non-autonomous
discrete KP hierarchy with a new technique. The second one is the
bilinearization obtained by introducing certain auxiliary $\tau$
function which has a similar structure to the ones that appeared in the
study of $R_{I}$ and $R_{II}$ biorthogonal rational
functions\cite{MT:RI,MT:ndToda}. The third bilinearization is through
the use of non-autonomous potential discrete KdV equation.

This article is organized as follows. In Section 2 we review the
bilinearization of the discrete KdV equation (\ref{eqn:dKdV}), and
discuss briefly why the similar calculation fails for the non-autonomous
case. In Section 3 we discuss the bilinearizations of
eq.(\ref{eqn:ndKdV}) and construct the Casorati determinant solution.
Finally, concluding remarks are given in Section 4.
\section{Bilinearization of the discrete KdV equation}
The discrete KdV equation (\ref{eqn:dKdV}) can be transformed to the
bilinear equation
 \begin{equation}
 \left(\frac{1}{a}+\frac{1}{b}\right) 
\tau_{n+1}^m\tau_{n-1}^{m+1}
- \left(\frac{1}{a}-\frac{1}{b}\right) 
\tau_{n-1}^{m}\tau_{n+1}^{m+1}
=\frac{2}{b}~\tau_n^m\tau_n^{m+1},
\label{bl:dKdV} 
\end{equation}
by the dependent variable transformation
\begin{equation}
 v_n^m = \frac{\tau_{n+1}^m\tau_{n}^{m+1}}{\tau_{n}^{m}\tau_{n+1}^{m+1}}.
\label{dep:dKdV}
\end{equation}
In fact, substituting eq.(\ref{dep:dKdV}) into eq.(\ref{eqn:dKdV}) we
have
\begin{align}
&   \left(\frac{1}{a}+\frac{1}{b}\right) 
\tau_{n+1}^m\tau_{n-1}^{m+1}\tau_{n}^{m+2}
-\left(\frac{1}{a}+\frac{1}{b}\right) 
\tau_{n-1}^{m+2}\tau_{n}^m\tau_{n+1}^{m+1}\nonumber\\
=&
 \left(\frac{1}{a}-\frac{1}{b}\right) 
\tau_{n-1}^{m}\tau_{n+1}^{m+1}\tau_{n}^{m+2}
-\left(\frac{1}{a}-\frac{1}{b}\right)
\tau_{n+1}^{m+2}\tau_{n}^m\tau_{n-1}^{m+1}.
\end{align}
Interchanging the second term of the left hand side and the first term
of the right hand side, and dividing the both sides by
$\tau_{n}^{m+2}\tau_{n}^m\tau_{n}^{m+1}$ we get

\begin{equation}
 \frac{
\left(\frac{1}{a}+\frac{1}{b}\right) 
\tau_{n+1}^m\tau_{n-1}^{m+1}
- \left(\frac{1}{a}-\frac{1}{b}\right) 
\tau_{n-1}^{m}\tau_{n+1}^{m+1}
}
{\tau_n^m\tau_n^{m+1}}
=
\frac{
\left(\frac{1}{a}+\frac{1}{b}\right) 
\tau_{n+1}^{m+1}\tau_{n-1}^{m+2}
-\left(\frac{1}{a}-\frac{1}{b}\right)
\tau_{n-1}^{m+1}\tau_{n+1}^{m+2}
}
{\tau_n^{m+1}\tau_n^{m+2}}. \label{mid_bl:dKdV}
\end{equation}
Equation (\ref{mid_bl:dKdV}) can be decoupled as
\begin{equation}
 \left(\frac{1}{a}+\frac{1}{b}\right) 
\tau_{n+1}^m\tau_{n-1}^{m+1}
- \left(\frac{1}{a}-\frac{1}{b}\right) 
\tau_{n-1}^{m}\tau_{n+1}^{m+1}
=\alpha(n)~\tau_n^m\tau_n^{m+1},
\end{equation}
since the right hand side of eq.(\ref{mid_bl:dKdV}) is obtained
from the left hand side by shifting $m$ to $m+1$ . 
Here $\alpha(n)$ is an arbitrary function in $n$, which can be 
absorbed by suitable gauge transformation on $\tau_n^m$. We obtain
eq.(\ref{bl:dKdV}) by choosing $\alpha(n)=\frac{2}{b}$ so that
$\tau_n^m=1$ is a solution.

The bilinear equation (\ref{bl:dKdV}) can be obtained by applying the
reduction to the Hirota-Miwa equation
\begin{align}
& a_1(a_2-a_3)~\tau(l_1+1,l_2,l_3)\tau(l_1,l_2+1,l_3+1)
+a_2(a_3-a_1)~\tau(l_1,l_2+1,l_3)\tau(l_1+1,l_2,l_3+1)\nonumber\\
+&a_3(a_1-a_2)~\tau(l_1,l_2,l_3+1)\tau(l_1+1,l_2+1,l_3)=0,\label{bl:dKP}
\end{align}
where $a_1$, $a_2$, $a_3$ are arbitrary constants. In fact, imposing the
condition
\begin{equation}
 \tau(l_1+1,l_2+1,l_3)\Bumpeq \tau(l_1,l_2,l_3),\label{reduction:dKdV}
\end{equation}
where $\Bumpeq$ means the equivalence up to gauge transformation, 
using eq.(\ref{reduction:dKdV}) to suppress the $l_1$ dependence and
putting $a_1=-a_2$, eq.(\ref{bl:dKP}) yields
\begin{equation}
\begin{split}
& -(a_2-a_3)~\tau(l_2-1,l_3)\tau(l_2+1,l_3+1)
+(a_2+a_3)~\tau(l_2+1,l_3)\tau(l_2-1,l_3+1)\\
&-2a_3~\tau(l_2,l_3+1)\tau(l_2,l_3)=0, 
\end{split}
\label{bl2:dKdV}
\end{equation}
which is equivalent to eq. (\ref{bl:dKdV}) with $l_2=n$, $l_3=m$,
$a_2=b$, $a_3=a$ and $\tau(l_2,l_3)=\tau_n^m$.

Now let us consider the non-autonomous case. We show that neither 
direct bilinearization nor reduction from the non-autonomous
Hirota-Miwa equation work
successfully for this case. First, substituting eq.(\ref{dep:dKdV}) into
eq.(\ref{eqn:ndKdV}) and doing the same calculation as above, we arrive at
the following equation
\begin{equation}
\begin{split}
& \frac{
\left(\frac{1}{a_m}+\frac{1}{b_n}\right) 
\tau_{n+1}^m\tau_{n-1}^{m+1}
- \left(\frac{1}{a_m}-\frac{1}{b_{n-1}}\right) 
\tau_{n-1}^{m}\tau_{n+1}^{m+1}
}
{\tau_n^m\tau_n^{m+1}}\\
=&
\frac{
\left(\frac{1}{a_{m+1}}+\frac{1}{b_{n-1}}\right) 
\tau_{n+1}^{m+1}\tau_{n-1}^{m+2}
-\left(\frac{1}{a_{m+1}}-\frac{1}{b_n}\right)
\tau_{n-1}^{m+1}\tau_{n+1}^{m+2}
}
{\tau_n^{m+1}\tau_n^{m+2}}, 
\end{split}
\end{equation}
which cannot be decoupled into the bilinear equation because of 
$n$ dependence of the coefficients. Therefore naive bilinearization
fails for the non-autonomous case. 

Secondly, let us consider the reduction from the non-autonomous
Hirota-Miwa equation\cite{WTS1,WTS2}
\begin{equation}
\begin{split}
& a_1(l_1)(a_2(l_2)-a_3(l_3))~\tau(l_1+1,l_2,l_3)\tau(l_1,l_2+1,l_3+1)\\
+&a_2(l_2)(a_3(l_3)-a_1(l_1))~\tau(l_1,l_2+1,l_3)\tau(l_1+1,l_2,l_3+1)\\
+&a_3(l_3)(a_1(l_1)-a_2(l_2))~\tau(l_1,l_2,l_3+1)\tau(l_1+1,l_2+1,l_3)=0,
\end{split}
\label{bl0:nHM} 
\end{equation}
where $a_i(l_i)$ $(i=1,2,3)$ are arbitrary functions. Imposing the
condition (\ref{reduction:dKdV}) on eq.(\ref{bl0:nHM}) and suppressing the
$l_1$-dependence, we obtain two
different bilinear equations
\begin{equation}
\begin{split}
& a_1(l_1)(a_2(l_2)-a_3(l_3))~\tau(l_2-1,l_3)\tau(l_2+1,l_3+1)\\
+&a_2(l_2)(a_3(l_3)-a_1(l_1))~\tau(l_2+1,l_3)\tau(l_2-1,l_3+1)\\
+&a_3(l_3)(a_1(l_1)-a_2(l_2))~\tau(l_2,l_3+1)\tau(l_2,l_3)=0,
\end{split}
\label{bl_mid1:nHM} 
\end{equation}
\begin{equation}
\begin{split}
& a_1(l_1-1)(a_2(l_2-1)-a_3(l_3))~\tau(l_2-1,l_3)\tau(l_2+1,l_3+1)\\
+&  a_2(l_2-1)(a_3(l_3)-a_1(l_1-1))~\tau(l_2+1,l_3)\tau(l_2-1,l_3+1)\\
+&a_3(l_3)(a_1(l_1-1)-a_2(l_2-1))~\tau(l_2,l_3+1)\tau(l_2,l_3)=0.
\end{split}\label{bl2:nHM}
\end{equation}
Since those two equations should be equivalent, the coefficients must satisfy
\begin{equation}
\begin{split}
&\frac{a_1(l_1)(a_2(l_2)-a_3(l_3))}{a_1(l_1-1)(a_2(l_2-1)-a_3(l_3))}
=\frac{a_2(l_2)(a_3(l_3)-a_1(l_1))}{a_2(l_2-1)(a_3(l_3)-a_1(l_1-1))}\\
=&\frac{a_3(l_3)(a_1(l_1)-a_2(l_2))}{a_3(l_3)(a_1(l_1-1)-a_2(l_2-1))},
\end{split}
\end{equation}
which yields
\begin{equation}
\begin{split}
& a_3(l_3)\left(\frac{1}{a_1(l_1)a_2(l_2-1)}-\frac{1}{a_1(l_1-1)a_2(l_2)}\right) \\
&+ \left(\frac{1}{a_1(l_1-1)}-\frac{1}{a_1(l_1)}\right)
-  \left(\frac{1}{a_2(l_2-1)}-\frac{1}{a_2(l_2)}\right)=0. 
\end{split}
\end{equation}
Since this should hold for any $a(l_3)$ for all $l_3$, we deduce that
$a(l_1)$ and $a(l_2)$ must be constants.  This implies that it is not
possible to impose the condition (\ref{reduction:dKdV}) on the
non-autonomous Hirota-Miwa equation (\ref{bl0:nHM}) consistently, unless
it is reduced to the autonomous case.
\section{Bilinearizations of the non-autonomous discrete KdV equation}
\subsection{Reduction from the discrete KP hierarchy}
The non-autonomous discrete KP hierarchy in the bilinear form is expressed as\cite{WTS1,OHTI:dKP}
\begin{equation}
 \left|\begin{array}{cccccc}
  1& a_{i_1}(l_{i_1})& a_{i_1}(l_{i_1})^2&\cdots &a_{i_1}(l_{i_1})^{m-2} &a_{i_1}(l_{i_1})^{m-2}\tau_{i_1}\tau_{\hat{i_1}} \\
  1& a_{i_2}(l_{i_2})& a_{i_2}(l_{i_2})^2&\cdots &a_{i_2}(l_{i_2})^{m-2} &a_{i_2}(l_{i_2})^{m-2}\tau_{i_2}\tau_{\hat{i_2}} \\
\vdots & \vdots & \vdots & & \vdots & \vdots\\
  1& a_{i_m}(l_{i_m})& a_{i_m}(l_{i_m})^2&\cdots &a_{i_m}(l_{i_m})^{m-2} &a_{i_m}(l_{i_m})^{m-2}\tau_{i_m}\tau_{\hat{i_m}} 
\end{array}\right|=0,
\label{bl:ndKP}
\end{equation}
where $\{i_1,\ldots,i_m\}\subset\{1,\ldots,n\}$, 
$\tau_{i_k}$ and $\tau_{\hat{i_k}}$ $(k=1,\ldots,m)$ are given by
\begin{align}
 \tau_{i_k} =& T_{i_k}\tau,\\
 \tau_{\hat{i_k}} =& T_{i_1}T_{i_2}\cdots T_{i_{k-1}}T_{i_{k+1}}\cdots T_{i_m}\tau,
\end{align}
respectively, $a_\nu(l_\nu)$ are arbitrary functions in $l_\nu$ for each
$\nu$, and $n$ and $m$ are arbitrary integers satisfying $n\geq m\geq 3$.  
Here $T_{i}$ is the shift operator of $l_i$ defined by
\begin{equation}
T_{i}\tau(l_1,l_2,\cdots,l_n)
=\tau(l_1,l_2,\cdots,l_{i-1},l_{i}+1,l_{i+1},\cdots,l_n).
\end{equation}
The simplest equation in
the hierarchy ($m=3$) is the non-autonomous Hirota-Miwa equation
\begin{equation}
\begin{split}
& a_i(l_i)(a_j(l_j)-a_k(l_k))~\tau(l_i+1,l_j,l_k)\tau(l_i,l_j+1,l_k+1)\\
+& a_j(l_j)(a_k(l_k)-a_i(l_i))~\tau(l_i,l_j+1,l_k)\tau(l_i+1,l_j,l_k+1)\\
+& a_k(l_k)(a_i(l_i)-a_j(l_j))~\tau(l_i,l_j,l_k+1)\tau(l_i+1,l_j+1,l_k)=0,
\end{split}
\label{bl:nHM} 
\end{equation}
where $\{i,j,k\}\subset\{1,\ldots,n\}$
and we suppressed other independent variables.
The Casorati determinant solution to the hierarchy can be written
as
\begin{equation}
 \tau(l_1,\cdots,l_n)=
\left|
\begin{array}{cccc}
 \varphi_1^{(s)}(l_1,\cdots,l_n)& \varphi_1^{(s+1)}(l_1,\cdots,l_n)
  &\cdots &  \varphi_1^{(s+N-1)}(l_1,\cdots,l_n)\\
 \varphi_2^{(s)}(l_1,\cdots,l_n)& \varphi_2^{(s+1)}(l_1,\cdots,l_n)
  &\cdots &  \varphi_2^{(s+N-1)}(l_1,\cdots,l_n)\\
\vdots &\vdots &\cdots &\vdots\\
 \varphi_N^{(s)}(l_1,\cdots,l_n)& \varphi_N^{(s+1)}(l_1,\cdots,l_n) &\cdots &  \varphi_N^{(s+N-1)}(l_1,\cdots,l_n)
\end{array}
\right|,
\end{equation}
where $\varphi_r^{(s)}$ ($r=1,\ldots,N$) satisfy the linear equations
\begin{equation}
 \frac{\varphi_r^{(s)}(l_1,\cdots,l_{\nu}+1,\cdots,l_n)
-\varphi_r^{(s)}(l_1,\cdots,l_{\nu},\cdots,l_n)}{a_\nu(l_\nu)}
=\varphi_r^{(s+1)}(l_1,\cdots,l_{\nu},\cdots,l_n),
\end{equation}
for $\nu=1,\ldots,n$. For example, the $N$-soliton solution is obtained
by choosing $\varphi_r^{(s)}$ as
\begin{equation}
 \varphi_r^{(s)}(l_1,\cdots,l_n) 
= \alpha_r p_r^s\prod_{\nu=1}^n \prod_{i=i_\nu}^{l_\nu-1}(1+a_\nu(i) p_r)
+ \beta_r q_r^s\prod_{\nu=1}^n \prod_{i=i_\nu}^{l_\nu-1}(1+a_\nu(i) q_r),
  \label{entry:ndKP}
\end{equation}
where $\alpha_r$, $\beta_r$, $p_r$, $q_r$  $(r=1,\ldots,N)$ are
arbitrary constants. 

Let us consider the reduction to the non-autonomous discrete KdV
equation. The key idea is to specialize some of the independent
variables to be autonomous by choosing the lattice intervals as
constants and use them as auxiliary variables. Then one can use the
autonomous variables for the reduction procedure to get the
non-autonomous discrete KdV equation.  We consider the four independent
variables $k=l_1$, $l=l_2$, $m=l_3$, $n=l_4$ with the lattice intervals
being $\delta=a_1(k)$, $\epsilon=a_2(l)$, $a_m=a_3(m)$, $b_n=a_4(n)$,
respectively. We note that $\delta$ and $\epsilon$ are constants,
namely, $k$ and $l$ are autonomous variables. Then we have the following
bilinear equations from eq.(\ref{bl:nHM}):
\begin{equation}
\begin{split}
& \delta(a_m-b_n)~\tau(k+1,l,m,n)\tau(k,l,m+1,n+1)\\
+&a_m(b_n-\delta)~\tau(k,l,m+1,n)\tau(k+1,l,m,n+1)\\
+&b_n(\delta-a_m)~\tau(k,l,m,n+1)\tau(k+1,l,m+1,n)=0,
\end{split}
\label{bl1:ndKP}
\end{equation}
\begin{equation}
\begin{split}
& \epsilon(a_m-b_n)~\tau(k,l+1,m,n)\tau(k,l,m+1,n+1)\\
+&a_m(b_n-\epsilon)~\tau(k,l,m+1,n)\tau(k,l+1,m,n+1)\\
+&b_n(\epsilon-a_m)~\tau(k,l,m,n+1)\tau(k,l+1,m+1,n)=0.
\end{split}\label{bl2:ndKP}
\end{equation}  
We impose the condition
\begin{equation}
 \tau(k+1,l+1,m,n)\Bumpeq \tau(k,l,m,n).\label{reduction:tau}
\end{equation}
This is achieved by imposing the condition on $\varphi_r^{(s)}$
($r=1,\ldots,N$) as
\begin{equation}
\varphi_r^{(s)}(k+1,l+1,m,n)\Bumpeq  \varphi_r^{(s)}(k,l,m,n) .\label{reduction:phi}
\end{equation}
For the case of soliton solutions, $\varphi_r^{(s)}(k,l,m,n)$ is
expressed as
\begin{align*}
 \varphi_r^{(s)}(k,l,m,n) =&\alpha_r
p_r^s(1+\delta p_r)^k(1+\epsilon p_r)^l \prod_{i=m_0}^{m-1}(1+a_ip_r)\prod_{j=n_0}^{n-1}(1+b_jp_r)\\
+&
\beta_r q_r^s(1+\delta q_r)^k(1+\epsilon q_r)^l\prod_{i=m_0}^{m-1}(1+a_iq_r)\prod_{j=n_0}^{n-1}(1+b_jq_r) .
\end{align*}
In order to satisfy eq.(\ref{reduction:phi}), one may take
\begin{equation}
 q_r = -p_r,\quad \delta=-\epsilon,
\end{equation}
so that
\begin{equation}
\begin{split}
& \varphi_r^{(s)}(k+1,l+1,m,n)=(1-\epsilon^2p_r^2)~\varphi_r^{(s)}(k,l,m,n),\\
& \tau(k+1,l+1,m,n)=\prod_{r=1}^{N}(1-\epsilon^2p_r^2)~\tau(k,l,m,n). 
\end{split}
\end{equation}
Then, suppressing the $k$-dependence by using eq.(\ref{reduction:tau}), the
bilinear equations (\ref{bl1:ndKP}) and (\ref{bl2:ndKP}) are reduced to
\begin{align*}
& -\epsilon(a_m-b_n)~\tau(l,m,n)\tau(l+1,m+1,n+1)
+a_m(b_n+\epsilon)~\tau(l+1,m+1,n)\tau(l,m,n+1)\\
&-b_n(\epsilon+a_m)~\tau(l+1,m,n+1)\tau(l,m+1,n)=0,\\[2mm]
& \epsilon(a_m-b_n)~\tau(l+1,m,n)\tau(l,m+1,n+1)
+a_m(b_n-\epsilon)~\tau(l,m+1,n)\tau(l+1,m,n+1)\\
&+b_n(\epsilon-a_m)~\tau(l,m,n+1)\tau(l+1,m+1,n)=0,
\end{align*}  
respectively. By putting
\begin{equation}
 \tau_n^m = \tau(l,m,n),\quad \sigma_n^m=\tau(l+1,m,n),
\end{equation}
the above bilinear equations are rewritten as
\begin{align}
& -\epsilon(a_m-b_n)~\tau_n^m\sigma_{n+1}^{m+1}
+a_m(b_n+\epsilon)~\tau_{n+1}^m\sigma_n^{m+1}
-b_n(\epsilon+a_m)~\tau_n^{m+1}\sigma_{n+1}^m=0, \label{bl1:ndKdV}\\[2mm]
& \epsilon(a_m-b_n)~\sigma_n^m\tau_{n+1}^{m+1}
+a_m(b_n-\epsilon)~\tau_n^{m+1}\sigma_{n+1}^m
+b_n(\epsilon-a_m)~\tau_{n+1}^m\sigma_n^{m+1}=0,\label{bl2:ndKdV}
\end{align}  
respectively. Equations (\ref{bl1:ndKdV}) and
(\ref{bl2:ndKdV}) can be regarded as a bilinearization of the
non-autonomous discrete KdV equation (\ref{eqn:ndKdV}). In fact,
introducing the variables $\Psi_n^m$ and $v_n^m$ by
\begin{subequations}
\begin{align}
& \Psi_n^m=\frac{\sigma_n^m}{\tau_n^m},\\
& v_n^m=\frac{\tau_{n+1}^m\tau_{n}^{m+1}}{\tau_{n}^{m}\tau_{n+1}^{m+1}}, \label{dep:ndKdV}
\end{align}
\end{subequations}
we obtain
\begin{align}
& \left(\frac{1}{b_n}-\frac{1}{a_m}\right)~\frac{1}{v_{n}^m}\Psi_{n+1}^{m+1}
-\left(\frac{1}{\epsilon}+\frac{1}{b_n}\right)~\Psi^{m+1}_n
+\left(\frac{1}{\epsilon}+\frac{1}{a_m}\right)~\Psi^m_{n+1}=0,\label{linear1:ndKdV}\\[2mm]
& \left(\frac{1}{b_n}-\frac{1}{a_m}\right)~\frac{1}{v_{n}^{m}}\Psi^m_n
+\left(\frac{1}{\epsilon}-\frac{1}{b_n}\right)~\Psi^m_{n+1}
+\left(\frac{1}{a_m}-\frac{1}{\epsilon}\right)~\Psi^{m+1}_n=0,\label{linear2:ndKdV}
\end{align}  
which are regarded as the auxiliary linear problem for the
non-autonomous discrete KdV equation. Eliminating $\Psi_m^n$
by considering the compatibility condition we obtain the non-autonomous
discrete KdV equation (\ref{eqn:ndKdV}). The $N$-soliton solution is
given by
\begin{equation}
 \tau_n^m=
\left|
\begin{array}{cccc}
\smallskip
 \varphi_1^{(s)}(m,n)& \varphi_1^{(s+1)}(m,n)  &\cdots &  \varphi_1^{(s+N-1)}(m,n)\\
\smallskip
 \varphi_2^{(s)}(m,n)& \varphi_2^{(s+1)}(m,n)  &\cdots &  \varphi_2^{(s+N-1)}(m,n)\\
\smallskip
\vdots &\vdots &\cdots &\vdots\\
 \varphi_N^{(s)}(m,n)& \varphi_N^{(s+1)}(m,n) &\cdots &  \varphi_N^{(s+N-1)}(m,n)
\end{array}
\right|,\label{tau:ndKdV}
\end{equation}
\begin{equation}
\varphi_r^{(s)}(m,n)=
\alpha_r p_r^s\prod_{i=m_0}^{m-1}(1+a_ip_r)\prod_{j=n_0}^{n-1}(1+b_jp_r)
+
\beta_r(-p_r)^s\prod_{i=m_0}^{m-1}(1-a_ip_r)\prod_{j=n_0}^{n-1}(1-b_jp_r) .\label{entry:tau}
\end{equation}
We remark that we obtain the non-autonomous potential discrete modified
KdV equation for $\Psi_n^m$ by eliminating $v_n^m$ from
eqs.(\ref{linear1:ndKdV}) and (\ref{linear2:ndKdV}).
\subsection{Alternate bilinearization}
There is another interesting bilinearization to the non-autonomous
discrete KdV equation (\ref{eqn:ndKdV}). Let us consider the following
bilinear equations,
\begin{align}
& b_{n}(a_{m-1}+a_m)~\kappa_{n+1}^m\tau_{n}^{m}
-a_{m-1}(a_m+b_{n})~\tau_{n+1}^{m-1}\tau_{n}^{m+1}
+a_m(a_{m-1}-b_{n})~\tau_{n+1}^{m+1}\tau_{n}^{m-1}=0,
\label{bl3:ndKdV}\\
& b_n(a_{m-1}-a_m)~\tau_n^{m+1}\tau_{n+1}^{m-1}
-a_m(a_{m-1}-b_n)~\tau_{n+1}^{m}\kappa_{n}^{m}
+a_{m-1}(a_m-b_n)~\tau_{n}^m\kappa_{n+1}^{m}=0.
\label{bl4:ndKdV}
\end{align}
We obtain eq.(\ref{eqn:ndKdV}) by introducing $v_n^m$ by
eq.(\ref{dep:ndKdV}) and eliminating $\kappa_n^m$.  The Casorati
determinant solution is given by
\begin{equation}
 \kappa_n^m=
\left|
\begin{array}{cccc}\smallskip
\psi_1^{(s)}(m,n)& \psi_1^{(s+1)}(m,n)&\cdots & \psi_1^{(s+N-1)}(m,n) \\
\smallskip
\psi_2^{(s)}(m,n)& \psi_2^{(s+1)}(m,n)&\cdots & \psi_2^{(s+N-1)}(m,n) \\
\smallskip
\vdots & \vdots &\cdots &\vdots\\
\psi_N^{(s)}(m,n)& \psi_N^{(s+1)}(m,n)&\cdots & \psi_N^{(s+N-1)}(m,n) 
\end{array}
\right|,
\end{equation}
where
\begin{equation}
\begin{split}
\psi_r^{(s)}(m,n)=&
\lambda_r p_r^s(1+a_mp_r)\prod_{i=m_0}^{m-2}(1+a_ip_r)\prod_{j=n_0}^{n-1}(1+b_jp_r)\\
+&
\mu_r(-p_r)^s(1-a_mp_r)\prod_{i=l_0}^{m-2}(1-b_ip_r)\prod_{j=n_0}^{n-1}(1-b_jp_r) ,
\end{split}\label{entry:kappa} 
\end{equation}
$\lambda_r$, $\mu_r$ are arbitrary constants ($r=1,\ldots,N$) and
$\tau_n^m$ is given by eqs.(\ref{tau:ndKdV}) and (\ref{entry:tau}).

We note that in the autonomous case, $\kappa_n^m$ reduces to $\tau_n^m$,
the bilinear equation (\ref{bl3:ndKdV}) yields eq.(\ref{bl:dKdV}), and
eq.(\ref{bl4:ndKdV}) becomes trivial, respectively. Secondly, because of
the symmetry with respect to $m$, $n$ in eqs.(\ref{eqn:ndKdV}) and
(\ref{dep:ndKdV}), the following bilinearization is also possible:
\begin{align}
& a_{m}(b_{n-1}+b_n)~\theta_n^{m+1}\tau_{n}^{m}
-b_{n-1}(b_n+a_{m})~\tau_{n-1}^{m+1}\tau_{n+1}^{m}
+b_n(b_{n-1}-a_{m})~\tau_{n+1}^{m+1}\tau_{n-1}^{m}=0,
\label{bl5:ndKdV}\\
& a_m(b_{n-1}-b_n)~\tau_{n+1}^{m}\tau_{n-1}^{m+1}
-b_n(b_{n-1}-a_m)~\tau_{n}^{m+1}\theta_{n}^{m}
+b_{n-1}(b_n-a_m)~\tau_{n}^m\theta_{n}^{m+1}=0,
\label{bl6:ndKdV}
\end{align}
whose solution is expressed as
\begin{equation}
\theta_n^m=
\left|
\begin{array}{cccc}\smallskip
\phi_1^{(s)}(m,n)& \phi_1^{(s+1)}(m,n)&\cdots & \phi_1^{(s+N-1)}(m,n) \\
\smallskip
\phi_2^{(s)}(m,n)& \phi_2^{(s+1)}(m,n)&\cdots & \phi_2^{(s+N-1)}(m,n) \\
\smallskip
\vdots & \vdots &\cdots &\vdots\\
\phi_N^{(s)}(m,n)& \phi_N^{(s+1)}(m,n)&\cdots & \phi_N^{(s+N-1)}(m,n) 
\end{array}
\right|,
\end{equation}
\begin{equation}
\begin{split}
\phi_r^{(s)}(m,n)=&
\lambda_r p_r^s(1+b_np_r)\prod_{i=m_0}^{m-1}(1+a_ip_r)\prod_{j=n_0}^{n-2}(1+b_jp_r)\\
+&
\mu_r(-p_r)^s(1-b_np_r)\prod_{i=l_0}^{m-1}(1-a_ip_r)\prod_{j=n_0}^{n-2}(1-b_jp_r) . 
\end{split}
\end{equation}
We also remark that the similar structure in the above auxiliary $\tau$
functions has appeared in the study of $R_I$ and $R_{II}$ biorthogonal
functions\cite{MT:RI,MT:ndToda}. Also, similar soliton type solution has been
constructed for the non-autonomous discrete-time Toda lattice
equation\cite{KM:ndToda}.

We can show that $\tau_n^m$ and $\kappa_n^m$ satisfy
eqs.(\ref{bl3:ndKdV}) and (\ref{bl4:ndKdV}) by the technique similar to
that was used in refs.\cite{MT:RI,MT:ndToda,KM:ndToda}. Namely, by
using the linear relations among $\varphi_r^{(s)}$ and $\psi_r^{(s)}$,
we first construct such difference formulas that express the
determinants whose columns are appropriately shifted by $\tau_n^m$ or
$\kappa_n^m$. Then eqs. (\ref{bl4:ndKdV}) and (\ref{bl5:ndKdV}) are
derived from Pl\"ucker relations which are quadratic identities of
determinants whose columns are shifted.

{}From eqs.(\ref{entry:tau}) and (\ref{entry:kappa}), we see that
$\varphi_r^{(s)}$ and $\psi_r^{(s)}$ satisfy
\begin{align}
& \varphi_r^{(s)}(m+1,n) - \varphi_r^{(s)}(m,n) = a_m~\varphi_r^{(s+1)}(m,n), \label{eqn:rec1}\\
& \varphi_r^{(s)}(m-1,n) + a_m~\varphi_r^{(s+1)}(m-1,n) = \psi_r^{(s)}(m,n), \label{eqn:rec2}\\
& \psi_r^{(s)}(m,n) - a_m~\psi_r^{(s+1)}(m,n) = (1-a_m^2p_r^2) ~\varphi_r^{(s)}(m-1,n),\label{eqn:rec3}\\
& \varphi_r^{(s)}(m,n+1) - \varphi_r^{(s)}(m,n) = b_n~\varphi_r^{(s+1)}(m,n). \label{eqn:rec6}
\end{align}
We introduce a notation 
\begin{equation}
 \tau_n^m 
=\left|~ 0,\ 1,\ \cdots,\ N-2,\ N-1~\right|,
\end{equation}
where ``$k$'' denotes the column vector
\begin{equation}
 k_{{m\atop n}} = \left(\begin{array}{c} \varphi_1^{(s+k)}(m,n)\\\vdots\\\varphi_N^{(s+k)}(m,n) \end{array}\right).
\end{equation}
Then the following difference formulas are derived from
eqs.(\ref{eqn:rec1})-(\ref{eqn:rec6}) by the similar calculations to
those given in ref.\cite{KM:ndToda}:
\begin{align}
-a_{m-1}\tau_n^{m-1}&=\left|~ 0, 1,\cdots,N-3, N-2,{N-2}_{m-1}~\right|,\label{eqn:dif1}\\
-b_{n-1}\tau_{n-1}^{m}&=\left|~ 0, 1,\cdots,N-3, N-2,{N-2}_{n-1}~\right|,\label{eqn:dif2}\\
A(m)^{-1}a_m~\tau_n^{m+1}&=  \left|~ 0, 1,\cdots,N-3, N-2,\widetilde{N-2}_{m+1}~\right|,\label{eqn:dif3}\\
-A(m)^{-1}(a_{m-1}+a_m)~\sigma_n^m &= \left|~ 0,1,\cdots,N-3,\widetilde{N-2}_{m+1}, N-2_{m-1}~\right|,\label{eqn:dif4}\\
-(a_{m-1}-b_{n-1})~\tau_{n-1}^{m-1}&=\left|~ 0, 1,\cdots,N-3, N-2_{n-1},{N-2}_{m-1}~\right|,\label{eqn:dif5}\\
A(m)^{-1}(a_{m}+b_{n-1})~\tau_{n-1}^{m+1}
&= \left|~ 0, 1,\cdots,N-3, N-2_{n-1},\widetilde{N-2}_{m+1}~\right|,\label{eqn:dif6}
\end{align}
where
\begin{equation}
 \widetilde{k}_{m+1}=
\left(
\begin{array}{c}
A_1(m)^{-1}~\varphi_1^{(s+k)}(m+1,n)\\\vdots\\A_N(m)^{-1}~\varphi_N^{(s+k)}(m+1,n)
\end{array}
\right),
\end{equation}
\begin{equation}
 A_r(m)=1-a_m^2p_r^2\quad (r=1,\ldots,N),\quad  A(m) = \prod_{r=1}^N (1-a_m^2p_r^2).
\end{equation}
We give the proof of the above formulas in the appendix. Applying
the the difference formulas to the Pl\"ucker relation
\begin{equation}
\begin{split}
0=& \left|~0,\cdots,N-3,N-2,N-2_{n-1}~\right|\times\left|~0,\cdots,N-3,N-2_{m-1},\widetilde{N-2}_{m+1}~\right|\\
-& \left|~0,\cdots,N-3,N-2,N-2_{m-1}~\right|\times\left|~0,\cdots,N-3,N-2_{n-1},\widetilde{N-2}_{m+1}~\right|\\
+& \left|~0,\cdots,N-3,N-2,\widetilde{N-2}_{m+1}~\right|\times\left|~0,\cdots,N-3,N-2_{n-1},N-2_{m-1}~\right|, 
\end{split}
\label{eqn:Pl1}
\end{equation}
we obtain the bilinear equation (\ref{bl4:ndKdV}). Equation
(\ref{bl5:ndKdV}) is derived by applying the difference formulas
(\ref{eqn:dif1}), (\ref{eqn:dif2}), (\ref{eqn:dif5}) and 
\begin{align}
-a_{m-2}~\sigma_{n}^{m-1}=&\left|~ 0, 1,\cdots,N-3, N-2,\widehat{N-2}_{m-1}~\right|,\label{eqn:dif10}\\
-(a_{m-2}-a_{m-1})~\tau_n^{m-2}=&\left|~ 0, 1,\cdots,N-3, N-2_{m-1},\widehat{N-2}_{m-1}~\right|,\label{eqn:dif11}\\
-(a_{m-2}-b_{n-1})~\sigma_{n-1}^{m-1}=&\left|~ 0, 1,\cdots,N-3, N-2_{n-1},\widehat{N-2}_{m-1}~\right|,\label{eqn:dif12}
\end{align}
where
\begin{equation}
\widehat{k}_{m-1}=\left(\begin{array}{c}\psi_1^{(s+k)}(m-1,n)\\\vdots\\\psi_N^{(s+k)}(m-1,n)\end{array}\right),
\end{equation}
to the Pl\"ucker relation,
\begin{equation}
\begin{split}
0=& \left|~0,\cdots,N-3,N-2,N-2_{n-1}~\right|\times\left|~0,\cdots,N-3,N-2_{m-1},\widehat{N-2}_{m-1}~\right|\\
-& \left|~0,\cdots,N-3,N-2,N-2_{m-1}~\right|\times\left|~0,\cdots,N-3,N-2_{n-1},\widehat{N-2}_{m-1}~\right|\\
+&
 \left|~0,\cdots,N-3,N-2,\widehat{N-2}_{m-1}~\right|\times\left|~0,\cdots,N-3,N-2_{n-1},N-2_{m-1}~\right|. 
\end{split}
\label{eqn:Pl2}
\end{equation}
\subsection{Reduction from the KP hierarchy through the potential form}
In this section, we consider the following difference equation
\begin{equation}
u_{n+1}^{m+1}- u_{n}^{m}   = \left(\frac{1}{a_m^2}-\frac{1}{b_{n}^2}\right)  \frac{1}{u_{n}^{m+1}-u_{n+1}^{m} },
\label{eqn:ndpKdV}
\end{equation}
which is closely related to eq.(\ref{eqn:ndKdV}) as 
\begin{equation}
 \left(\frac{1}{a_m}-\frac{1}{b_{n}}\right)  \frac{1}{v_{n}^{m}}
= u_{n+1}^{m} - u_{n}^{m+1}.\label{eqn:ndpKdV_u}
\end{equation}
The autonomous version of eq.(\ref{eqn:ndpKdV}) is known as the
potential discrete KdV equation\cite{Nijhoff-Capel:dKdV}. We call eq.(\ref{eqn:ndpKdV}) the
non-autonomous potential discrete KdV equation.
Casorati determinant solution to eq.(\ref{eqn:ndpKdV}) is given by
\begin{equation}
 u_n^m =
  \frac{\rho_n^m}{\tau_{n}^m}-\sum_{i=m_0}^{m-1}\frac{1}{a_i}-\sum_{j=n_0}^{n-1}\frac{1}{b_j},
\label{eqn:ndpKdV_dep}
\end{equation}
where
\begin{equation}
 \rho_{n}^{m} = 
\left|
\begin{array}{cccc}
 \varphi_1^{(s)}(m,n)&\cdots & \varphi_1^{(s+N-2)}(m,n)& \varphi_1^{(s+N)}(m,n) \\
 \varphi_2^{(s)}(m,n)&\cdots & \varphi_2^{(s+N-2)}(m,n)& \varphi_2^{(s+N)}(m,n) \\
\vdots & \vdots &\cdots         &\vdots\\
 \varphi_N^{(s)}(m,n)&\cdots & \varphi_N^{(s+N-2)}(m,n)& \varphi_N^{(s+N)}(m,n) 
\end{array}
\right|,\label{eqn:ndpKdV_rho}
\end{equation}
and $\varphi_r^{(s)}(m,n)$ ($r=1,\ldots,N$), $\tau_n^m$ are defined by
eqs.(\ref{entry:tau}) and (\ref{tau:ndKdV}), respectively. 

Equation (\ref{eqn:ndpKdV}) is derived from the following bilinear
equations for $\rho_n^m$ and $\tau_n^m$
\begin{align}
& \rho_{n}^{m+1}\tau_{n+1}^{m} -\rho_{n+1}^{m}\tau_{n}^{m+1}=
\left(\frac{1}{a_m}-\frac{1}{b_{n}}\right)~\left(\tau_{n}^{m+1}\tau_{n+1}^{m}-\tau_{n+1}^{m+1}\tau_{n}^{m}\right),
\label{eqn:ndpKdV_bl1}\\ 
& \rho_{n+1}^{m+1}\tau_{n}^{m} -\rho_{n}^{m}\tau_{n+1}^{m+1}=
\left(\frac{1}{a_m}+\frac{1}{b_{n}}\right)~\left(\tau_{n+1}^{m+1}\tau_{n}^{m}-\tau_n^{m+1}\tau_{n+1}^m\right),
\label{eqn:ndpKdV_bl2} 
\end{align}
through the dependent variable transformation (\ref{eqn:ndpKdV_dep}). In
particular, eq.(\ref{eqn:ndpKdV_u}) also follows from eq.(\ref{eqn:ndpKdV_bl1}). Therefore, we may regard
eqs.(\ref{eqn:ndpKdV_bl1}) and (\ref{eqn:ndpKdV_bl2}) as yet another
bilinearization of the non-autonomous discrete KdV equation
(\ref{eqn:ndKdV}).

We can show that $\rho_n^m$ and $\tau_n^m$ satisfy
eqs. (\ref{eqn:ndpKdV_bl1}) and (\ref{eqn:ndpKdV_bl2})
as follows. Applying the difference formulas (\ref{eqn:dif1}),
(\ref{eqn:dif2}), (\ref{eqn:dif5}) and
\begin{align}
  -(a_{m-1}\rho_{n}^{m-1}+\tau_{n}^{m-1})&=|~0,1,\cdots,N-3,N-1,N-2_{m-1}~|,\label{eqn:dif13}\\
  -(b_{n-1}\rho_{n-1}^{m}+\tau_{n-1}^{m})&=|~0,1,\cdots,N-3,N-1,N-2_{n-1}~|, \label{eqn:dif14}
\end{align}
to the Pl\"ucker relation
\begin{equation}
\begin{split}
0&= \left|~0,\cdots,N-3,N-2,N-1~\right|\times\left|~0,\cdots,N-3,N-2_{n-1},N-2_{m-1}~\right|\\
&- \left|~0,\cdots,N-3,N-2,N-2_{n-1}~\right|\times\left|~0,\cdots,N-3,N-1,N-2_{m-1}~\right|\\
&+ \left|~0,\cdots,N-3,N-2,N-2_{m-1}~\right|\times\left|~0,\cdots,N-3,N-1,N-2_{n-1}~\right|,
\end{split}
\label{eqn:ndpKdV_Pl1}  
\end{equation}
we have eq. (\ref{eqn:ndpKdV_bl1}). Similarly, we obtain
eq.(\ref{eqn:ndpKdV_bl2}) by applying the formulas (\ref{eqn:dif2}),
(\ref{eqn:dif3}), (\ref{eqn:dif6}), (\ref{eqn:dif14}) and
\begin{equation}
 A(m)^{-1}(a_m\rho_n^{m+1}-\tau_n^{m+1})=|~ 0, \cdots, N-3, N-1,\widetilde{N-2}_{m+1}~|,
\end{equation}
to the Pl\"ucker relation
\begin{equation}
\begin{split}
0&= \left|~0,\cdots,N-3,N-2,N-1~\right|\times\left|~0,\cdots,N-3,N-2_{n-1},\widetilde{N-2}_{m+1}~\right|\\
&- \left|~0,\cdots,N-3,N-2,N-2_{n-1}~\right|\times\left|~0,\cdots,N-3,N-1,\widetilde{N-2}_{m+1}~\right|\\
&+ \left|~0,\cdots,N-3,N-2,\widetilde{N-2}_{m+1}~\right|\times\left|~0,\cdots,N-3,N-1,N-2_{n-1}~\right| .
\end{split}
\label{eqn:ndpKdV_Pl2}  
\end{equation}
We finally remark that if we introduce the continuous
independent variables $t_1$, $t_3$, $\cdots$ through $\varphi_r^{(s)}(m,n)$
as
\begin{equation}
\begin{split}
 \varphi_r^{(s)}(m,n) =& \alpha_r p_r^s\prod_{i=m_0}^{m-1}(1+a_ip_r)\prod_{j=n_0}^{n-1}(1+b_jp_r)~e^{p_rt_1+p_r^3t_3+\cdots}\\
+&
\beta_r(-p_r)^s\prod_{i=m_0}^{m-1}(1-a_ip_r)\prod_{j=n_0}^{n-1}(1-b_jp_r)~e^{-p_rt_1-p_r^3t_3+\cdots}, 
\end{split}
\end{equation}
then $\tau_n^m$ becomes the $\tau$ function of the KdV hierarchy.
In this case, $\rho_n^m$ and $u_n^m$ can be expressed as
\begin{equation}
\rho_n^m = \frac{\partial \tau_n^m}{\partial  t_1},\quad
u_n^m = \frac{\partial }{\partial t_1}\log \tau_n^m,
\end{equation}
respectively,  and $u_n^m$ satisfies the potential KdV equation
\begin{equation}
 \frac{\partial u_n^m}{\partial t_3} - \frac{3}{2}\left(\frac{\partial u_n^m}{\partial t_1}\right)^2
-\frac{1}{4}\frac{\partial^3 u_n^m}{\partial t_1^3}=0.
\end{equation}
This is consistent with the fact that (autonomous version of) eq.(\ref{eqn:ndpKdV}) is derived
as the B\"acklund transformation of the potential KdV equation\cite{Nijhoff-Capel:dKdV}.
\section{Concluding remarks}
In this article, we have considered the bilinearization of the
non-autonomous discrete KdV equation and constructed Casorati
determinant solution. We have presented three different
bilinearizations, each of which has different origin. Although we have
constructed only Casorati determinant solution, namely, soliton type
solution, it might not be difficult to discuss other types of solutions,
such as rational solutions or periodic solutions, based on the bilinear
equations that have been obtained in this article. Also, we expect that
other non-autonomous discrete integrable systems on two-dimensional
lattice can be investigated in similar manner. 

As was mentioned in Section 3.2, the $\tau$ functions in the second
bilinearization resemble those in the theory of $R_{I}$ and $R_{II}$
biorthogonal functions, but the explicit relation is not clear yet. It
might be an intriguing problem to study underlying structure of
the second bilinearization.  

Finally, recently Takahashi and Hirota have succeeded in constructing the
soliton solutions of the ultradiscrete KdV equation in permanent
form\cite{TH:uKdV_permanent}. It might be an interesting problem to investigate
the permanent type solutions for the non-autonomous case.
\section*{Acknowledgments}
The authors would like to express their sincere gratitude to Professor
N. Matsuura for stimulating discussions and fruitful informations which
motivated this work.  They are also grateful to Professor A. Nakayashiki
for valuable discussions and useful comments.
\appendix
\section{Proof of difference formulas} In the appendix, we give the
proof of the difference formulas of $\tau$ functions which have been
used in the derivation of bilinear equations from the Pl\"ucker
relations for completeness.  For later convenience, we first prepare the
following two equations for $\varphi_r^{(s)}$ and $\psi_r^{(s)}$ which
are derived from eqs.(\ref{eqn:rec1})-(\ref{eqn:rec3}):
\begin{align}
& \psi_r^{(s)}(m,n) + a_{m-1}~\psi_r^{(s+1)}(m,n) = \varphi_r^{(s)}(m+1,n), \label{eqn:rec4}\\
& \varphi_r^{(s)}(m+1,n) - a_m~\varphi_r^{(s+1)}(m+1,n) = A_r(m)~\varphi_r^{(s)}(m,n).\label{eqn:rec5}
\end{align}

\paragraph{Equations (\ref{eqn:dif1}) and (\ref{eqn:dif2})}
We have
\begin{equation}
 \tau_n^{m-1}=\left|~0_{m-1},1_{m-1},\cdots,N-2_{m-1},N-1_{m-1}\right|.
\end{equation}
Adding the $(i+1)$-th column multiplied by $a_{m-1}$ to the $i$-th column for
$i=1,2,\ldots,N-1$ and using eq.(\ref{eqn:rec1}), we have
\begin{equation}
 \tau_n^{m-1}= \left|~0,1,\cdots,N-2,N-1_{m-1}\right|.
\end{equation}
Multiplying $a_{m-1}$ to the $N$-th column and using eq.(\ref{eqn:rec1}) we obtain
\begin{align*}
a_{m-1}~\tau_n^{m-1}=&\left|~0,1,\cdots,N-2,a_{m-1}\times(N-1)_{m-1}\right| \\
=&\left|~0,1,\cdots,N-2,(N-2)_m-(N-2)_{m-1}\right| \\
=&-\left|~0,1,\cdots,N-2,N-2_{m-1}\right| ,
\end{align*}
which is eq.(\ref{eqn:dif1}). Equation (\ref{eqn:dif2}) can be shown in
a similar manner by shifting $n$ and using eq.(\ref{eqn:rec6}).
\paragraph{Equation (\ref{eqn:dif3})} We have
\begin{equation}
 \tau_n^{m+1}=\left|~ 0_{m+1},\ 1_{m+1},\ \cdots,\ N-2_{m+1},\ N-1_{m+1}~\right|.
\end{equation}
Adding the $(i+1)$-th column multiplied by $-a_{m}$ to the $i$-th column for
$i=1,2,\ldots,N-1$ and using eq.(\ref{eqn:rec5}), we have
\begin{displaymath}
 \tau_n^{m+1} 
= \left|~ \overline{0},\ \overline{1},\ \cdots,\ \overline{N-2},\ N-1_{m+1}~\right|,
\quad \overline{k}=
\left(\begin{array}{c}A_1(m)~\varphi_1^{(s)}(m,n)\\\vdots\\A_N(m)~\varphi_N^{(s)}(m,n)\end{array}\right).
\end{displaymath}
Multiplying $a_m$ to the $N$-th column and using eq.(\ref{eqn:rec5}) we have
\begin{align*}
  a_m~\tau_n^{m+1}=&\left|~\overline{0},\overline{1},\cdots,\overline{N-2},a_m\times(N-1)_{m+1}~\right|\\
=&\left|~\overline{0},\overline{1},\cdots,\overline{N-2},N-2_{m+1}~\right|\\
=&A(m)\times \left|~ 0,1,\cdots,N-2,\widetilde{N-2}_{m+1}~\right|,
\end{align*}
which is eq.(\ref{eqn:dif3}).
\paragraph{Equation (\ref{eqn:dif4})}
We have
\begin{equation}
  \sigma_n^m = \left|~\widehat{0},\widehat{1},\cdots,\widehat{N-1}~\right|,
\end{equation}
which is rewritten by using eq.(\ref{eqn:rec4}) from the first column to
the $N$-th column as
\begin{equation}
 \sigma_n^m =\left|~0_{m+1},1_{m+1},\cdots,N-2_{m+1},\widehat{N-1}_m~\right|. \label{eqn:dif41}
\end{equation}
Now, from eqs.(\ref{eqn:rec3}) and (\ref{eqn:rec4}) we have
\begin{equation}
 A_r(m)~\varphi_r^{(s)}(m-1,n) = - (a_m+a_{m-1})~\psi_r^{(s+1)}(m,n) + \varphi_r^{(s)}(m+1,n).
\label{eqn:rec11}
\end{equation}
Applying eq. (\ref{eqn:rec11}) to the $N$-th column of the right hand side
of eq.(\ref{eqn:dif41}) we obtain
\begin{align*}
(a_m+a_{m-1})~\sigma_n^m =&-\left|~0_{m+1},\cdots,N-2_{m+1},\ \overline{N-2}_{m-1}~\right|\\
=&-\left|~\overline{0}_m,\cdots,\overline{N-3}_m,\ N-2_{m+1},\ \overline{N-2}_{m-1}~\right|
\\
=&-A(m)\times
\left|~0_{m},\cdots,N-3_m,\widetilde{N-2}_{m+1},N-2_{m-1}~\right|,
\end{align*}
which is eq.(\ref{eqn:dif4}).
\paragraph{Equation (\ref{eqn:dif5})}
Shifting $n$ in eq.(\ref{eqn:dif1}) we have
\begin{equation}
-a_{m-1}\tau_{n-1}^{m-1}=\left|~ 0_{n-1}, 1_{n-1},\cdots,N-3_{n-1},
 N-2_{n-1},{N-2}_{{m-1}\atop{n-1}}~\right|. \label{eqn:dif51}
\end{equation}
Eliminating $\varphi_r^{(s+1)}(m-1,n-1)$ from
eqs.(\ref{eqn:rec1})$_{{m-1}\atop{n-1}}$ and
(\ref{eqn:rec6})$_{{m-1}\atop{n-1}}$ we get
\begin{equation}
 (a_{m-1}-b_{n-1})~\varphi_r^{(s)}(m-1,n-1)=a_{m-1}\varphi_r^{(s)}(m-1,n)
-b_{n-1}\varphi_r^{(s)}(m,n-1).\label{eqn:rec12}
\end{equation}
Using eq.(\ref{eqn:rec12}) to the $N$-th column of the right hand side
of eq.(\ref{eqn:dif51}), we have
\begin{displaymath}
 -a_{m-1} (a_{m-1}-b_{n-1})~\tau_{n-1}^{m-1}
=a_{m-1}\left|~ 0_{n-1}, 1_{n-1},\cdots,N-3_{n-1}, N-2_{n-1},{N-2}_{m-1}~\right| .
\end{displaymath}
Adding $(i+1)$-th column multiplied by $a_{m-1}$ to $i$-th column for
$i=1,2,\ldots,N-1$ and using eq.(\ref{eqn:rec1}), we obtain
\begin{displaymath}
  -(a_{m-1}-b_{n-1})~\tau_{n-1}^{m-1}=\left|~ 0, 1,\cdots,N-3, N-2_{n-1},{N-2}_{m-1}~\right|,
\end{displaymath}
which is eq.(\ref{eqn:dif5}).
\paragraph{Equation (\ref{eqn:dif6})}
Shifting $n$ in eq.(\ref{eqn:dif3}) we have
\begin{align*}
A(m)^{-1}a_m~\tau_{n-1}^{m+1}
=&\left|~ 0_{n-1}, 1_{n-1},\cdots,N-3_{n-1}, N-2_{n-1},
\widetilde{N-2}_{{m+1}\atop{n-1}}~\right|\\
=&\left|~ 0, 1,\cdots,N-3, N-2_{n-1},\widetilde{N-2}_{{m+1}\atop{n-1}}~\right|,
\end{align*}
or
\begin{equation}
 a_m~\tau_{n-1}^{m+1}=
\left|~ \overline{0}, \overline{1},\cdots,\overline{N-3}, \overline{N-2}_{n-1},N-2_{{m+1}\atop{n-1}}~\right|.
\label{eqn:dif61}
\end{equation}
Eliminating  $\varphi_r^{(s+1)}(m+1,n-1)$ from
eqs.(\ref{eqn:rec5})$_{n-1}$ and (\ref{eqn:rec6})$_{{m+1}\atop{n-1}}$ we
get 
\begin{equation}
b_{n-1} A_r(m)~\varphi_r^{(s)}(m,n-1)+a_m~\varphi_r^{(s)}(m+1,n)
=(a_m+b_{n-1})~\varphi_r^{(s)}(m+1,n-1).\label{eqn:rec13}
\end{equation}
Applying eq.(\ref{eqn:rec13}) to the $N$-th column of the right hand side
of eq.(\ref{eqn:dif61}), we have
\begin{displaymath}
  a_m(a_m+b_{n-1})~\tau_{n-1}^{m+1}
=a_m\times
\left|~ \overline{0}, \cdots,\overline{N-3}, \overline{N-2}_{n-1},\ N-2_{m+1}~\right|,
\end{displaymath}
which yields eq.(\ref{eqn:dif6})
\begin{displaymath}
(a_m+b_{n-1})~\tau_{n-1}^{m+1}
=A(m)\times
\left|~ 0, \cdots,N-3, N-2_{n-1},\widetilde{N-2}_{m+1}~\right|.
\end{displaymath}
We omit the proof of other difference formulas, since they can be proved
in a similar manner.

\end{document}